\documentclass[aps,pre,amsmath,amssymb,nofootinbib]{revtex4-1}
\usepackage{graphicx}
\usepackage[]{epsfig}
\usepackage{amsmath,pgfmath,pgffor}
\usepackage[usenames, dvipsnames]{color}
\usepackage{indentfirst}
\usepackage[utf8]{inputenc}
\usepackage{hyperref}
\usepackage{subfigure}

\usepackage{hyperref}

\usepackage{color}

\hypersetup{%
    pdfborder = {0 0 0},
    colorlinks,
    citecolor=blue,
    filecolor=Darkgreen,
    linkcolor=black,
}

\begin{document}

\title{Higher order corrections to the effective potential close to the jamming transition\\
 in the perceptron model}

\author{Ada Altieri} 
\email{altieri.ada@gmail.com} 
\affiliation{Dipartimento di Fisica, Sapienza Universit\`a di Roma, Piazzale A. Moro 2, I-00185, Rome, Italy}
\affiliation{LPTMS, CNRS, Univ. Paris-Sud, Universit\'e Paris-Saclay, 91405 Orsay, France}

\begin{abstract}
We analyze the perceptron model performing a Plefka-like expansion of the free energy. This model falls in the same universality class as hard spheres near jamming, allowing to get exact predictions in high dimensions for more complex systems.
Our method enables to define an effective potential (or TAP free energy), namely a coarse-grained functional depending on the contact forces and the effective gaps between the particles. The derivation is performed up to the third order, with a particular emphasis on the role of third order corrections to the TAP free energy. These corrections, irrelevant in a mean-field framework in the thermodynamic limit, might instead play a fundamental role when considering finite-size effects. 
We also study the typical behavior of the forces and we show that two kinds of corrections can occur. The first contribution arises since the system is analyzed at a finite distance from jamming, while the second one is due to finite-size corrections. In our analysis, third order contributions vanish in the jamming limit, both for the potential and the generalized forces, in agreement with the argument proposed by Wyart and coworkers invoking isostaticity. 
Finally, we analyze the scalings emerging close to the jamming line, which define a crossover regime connecting the control parameters of the model to an effective temperature. 

\end{abstract}

\maketitle

\section{Introduction}

The anomalous properties of low-temperature structural glasses have been the object of intense studies for decades. By analyzing a vast class of materials with only repulsive contact interactions - for instance emulsions, hard-sphere suspensions, granular media - a new kind of transition has been detected \cite{Liu-Nagel, Durian-Weitz, OHern, Silbert-Liu, Sperl-Majmudar}, \emph{the jamming transition}, consisting in the passage from a fluid phase to a regime characterized by a stiff arrangement of particles unable to move and flow.
While the glass transition is generated by a rapid cooling down of the liquid in order to avoid crystallization, the jamming transition is induced by an increasing density protocol in the zero-temperature limit. This defines a purely geometric problem where thermal energy is not relevant in determining or facilitating the transition. Anyway, the analytical investigation of the jamming transition turns out to be a very challenging issue, both in a mean-field perspective and in finite dimension. 

Very recently a breakthrough has been achieved in the context of hard-sphere systems in the limit of infinite space dimensions \cite{CKPUZ,KPZ,KPUZ_II,CKPUZ_II, Biroli-Urbani, BU}.
In this context, the possibility of establishing a unifying framework for jamming, irrespective of microscopic details and the specific numerical setup, looks very intriguing. Indeed, several properties of the jamming transition - as the emergence of a power-law behavior in the distribution of the forces and gaps between the particles, the nature and the shape of vibrational modes \cite{OHern, Silbert-Liu, Wyart1} - turn out to be independent of the protocol.

The underlying idea of a sort of universal behavior goes beyond mechanical considerations, involving a broader class of systems, known as continuous constraint satisfaction problems (CSP), where a set of constraints is imposed on a set of continuous variables. Similarly, in a jammed system the particle motion is hindered by neighboring particles, which induce geometrical and mechanical constraints in terms of force and torque balance. 
The connection between jammed systems and the CSP paradigm has been proposed in several works \cite{KK, ZK, Mari-Kurchan-Krzakala}. 
However, further developments in this field have been made possible once people realized that sphere systems in high dimension belong to the same universality class as a simplified model, the perceptron, according to a new interpretation proposed by Franz and Parisi \cite{SimplestJamming}. 

The perceptron model has been exploited as a linear signal classifier in computer science for many years \cite{Gardner-Derrida, Gardner}. It is nevertheless proposed here in a modified version \cite{FPUZ, AFP, FPSUZ}, with a particular emphasis on a regime that gives rise to non-convex properties in the space of solutions. We shall clarify this point in more details later.

The Franz-Parisi model is a remarkable starting point for studying jamming in high dimensions. It essentially consists in $M$ obstacles randomly distributed over a spherical surface in $N$ dimensions, with radius $\sqrt{N}$. The positions of the particles must satisfy specific constraints, which affect the general properties of the model and the energy value. For each violated constraint there is an associated energy cost to pay.

The Hamiltonian of the model depends on $M=\alpha N$ random {\it gaps} $h_\mu(\vec{x})$ (where $\mu=1,...,M)$ via a soft-constraint interaction:
\begin{equation}
\mathcal{H}[\vec{x}]=\frac{1}{2} \sum_{\mu=1}^{M} h_{\mu}^2(\vec{x}) \theta(-h_\mu(\vec{x})) \ ,
\label{Hamiltonians}
\end{equation} 
where $\theta(x)$ is the Heaviside function.
The gaps are functions of the system configuration $\vec{x}=\lbrace x_1,...,x_N \rbrace$, defined on a $N$-dimensional hypersphere, \emph{i.e.} $\sum \limits_{i=1}^{N}x_i^2=N$. They satisfy the following relation:
\begin{equation}
h_{\mu}(\vec{x})=\sum_{i=1}^{N} \frac{\xi_i^{\mu}  x_i}{\sqrt{N}} -\sigma \ ,
\label{Gap_x}
\end{equation}
corresponding to the scalar product between the random obstacles $\xi_i^{\mu}$  ($i=1,...,N$ and $\mu=1,...,M$) and the reference particle position. The components $\xi^{\mu}_i$, which play the role of quenched disorder, are i.i.d random variables distributed according to a normal distribution $\mathcal{N}(0,1)$. 
By conveniently varying the two tunable parameters, $\sigma$ and $\alpha=M/N$, one might explore different regions of the phase diagram. In particular, the system might undergo a critical transition determining the passage from a satisfiable region, SAT phase (where at least one configuration $\vec{x}$ satisfies simultaneously all the constraints), to an unsatisfiable one, UNSAT phase, with all the constraints not verified at the same time. The first one corresponds to a hard-sphere (HS) regime defined by a zero energy manifold, whereas the second scenario can be mapped to a soft-sphere (SS) problem described by a harmonic potential in $h_\mu(\vec{x})$ (see Eq.(\ref{Hamiltonians})). Physically, this SAT/UNSAT transition coincides with the jamming transition in correspondence of which the volume of the space of solutions satisfying the given assignments continuously shrinks to zero. 

Depending on the positive or negative value of the control parameter $\sigma$, two different situations can occur. For positive $\sigma$ the model defines the usual perceptron classifier, which gives rise to a convex optimization problem, whereas for negative $\sigma$ the space of solutions is no longer convex, loosing its ergodicity properties and inducing new interesting features. Indeed, it can be regarded as the problem of a single dynamical sphere in a background of \emph{quenched} obstacles $\xi^{\mu}$, often called \emph{patterns} in neural network notation. This is exactly the regime we are interested in, allowing to map the problem to more complex settings. Essentially, this model bridges the gap between generic CSPs and disordered sphere packings, thanks to the fact that it displays a similar phase diagram to the one of hard spheres in high dimensions and a jamming transition belonging to the same universality class. 
\begin{figure}[h]
\centering
\includegraphics[height=2.in]{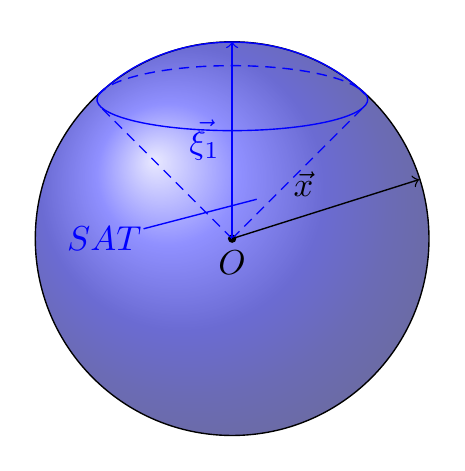}
\caption{Schematic illustration of the perceptron model in the convex case, \emph{i.e.} for $\sigma >0$, with only one constraint. The scalar product between $\vec{x}$ and $\vec{\xi}_1$ should be bigger than a threshold, hence excluding all the solutions in the region below the dashed cone. The space of solution is now convex. With more constraints the space of solutions reduces, remaining nevertheless always convex. Both the reference particle and the obstacles live on a $N$-dimensional sphere of radius $\sqrt{N}$. For $\sigma < 0 $ the picture would be different, given by the intersection of non-convex domains.}
\end{figure}

In this paper we aim to determine an effective thermodynamic potential that properly describes the perceptron model. It turns out to be a central issue especially in the SAT phase, where the energy manifold is flat and unseemly to analyze small harmonic fluctuations around the metastable states of the systems, \emph{i.e.} the minima of a suitable functional. Therefore, we need to introduce coarse grained variables and to formulate a systematic approach to define a free energy landscape. At the end it will be possible to perform a detailed analysis of the most relevant features of the SAT phase and the jamming regime.

In \cite{AFP} the definition of an effective potential as a function of local order parameters, namely of both the average particle positions and the \emph{generalized forces}, has been proposed. Generalized in the sense that they result from the differentiation of the potential with respect to the effective gaps rather than to the particle positions.
Our computation is focused on a formal coupling expansion of the free energy, which actually coincides with a perturbative diagrammatic expansion in $1/N$, valid both in the liquid phase and in the low-temperature regime. In our previous work only the first two moments of the expansion have been taken into account within a mean-field-like picture. The main goals of this paper are instead: i) the computation of third order corrections to the effective potential, which might also capture important features in the physics of low-temperature glasses in finite dimensions; ii) and the estimation of subleading contributions to the generalized forces.

The paper is organized as follows: in Section II we introduce the mathematical details which allow defining the model and we briefly summarize the main steps for the derivation of the effective potential in the SAT phase. In Section III we compute third-order corrections and we explain why their contribution reasonably vanishes at jamming. Once defined a coarse-grained free energy as a function of both particle positions and contact forces, in Section IV we evaluate the leading and subleading behavior of the forces. We highlight the emergent contributions of terms that can be incorporated in a generic scaling function, which differentiates between the jamming limit and the more general case. Two kinds of corrections in the force expression are expected: one due to the subleading terms in the asymptotic expansion of the potential near the jamming line, visible even in a mean-field framework, and another due to finite-size corrections in ordinary systems. Finally, in Section V we propose a scaling argument based on the full RSB ansatz, which provides a fairly accurate estimate of a crossover regime as a function of an effective temperature.

\vspace{0.5cm}

\section{TAP free energy for the negative perceptron}
In \cite{AFP} the effective thermodynamic potential $\Gamma(\vec{m}, \vec{f})$, as a function of both the particle positions and the forces, has been derived for the perceptron model. In the following we shall give a hint of the analytical scheme to follow in order to calculate the effective potential in high-dimensional systems. We shall exploit a small coupling expansion according to the formalism first proposed by Plefka \cite{Plefka} and later reformulated by Georges and Yedidia \cite{Georges-Yedidia}. Thanks to the fully-connected structure of the model, the calculations lead to a simplified derivation and to a reasonable truncated expansion after a finite number of terms. This approach provides a non-convex free energy functional that nevertheless gives access to the metastable states of the system. 

Given the definition of the gaps in Eq. (\ref{Gap_x}), we enforce that $h_\mu=h_\mu(x)$ in the partition function via $M$ auxiliary variables $i\hat{h}_\mu$ conjugated to the gaps. The average values of the forces and the positions, which the free energy functional actually depends on, are enforced via the Lagrange multipliers $u_i$ and $v_\mu$.
\begin{eqnarray}
  \label{eq:due}
  e^{-\Gamma(\vec{m},\vec{f})}=\int d\vec{x} d{\vec{h}} d \vec{\hat{h}} \; e^{-\beta H[\vec{h}]+{\sum_i (x_i-m_i)u_i +\sum_\mu (i \hat{h}_\mu-f_\mu) v_\mu +\sum_{\mu} i \hat{h}_\mu( h_\mu(x)-h_\mu)}
}=e^{J(\vec{u}, \vec{v}) -{\vec{m} \cdot \vec{u}} -\vec{ f}  \cdot \vec{v}} \ ,
\end{eqnarray}
with $\frac{\partial J}{\partial u_i}=\frac{\partial J}{\partial v_\mu}=0$, $ \forall i,\mu$. The functional $\Gamma(\vec{m}, \vec{f})$ reads:
\begin{equation}
\Gamma(\vec{m},\vec{f})= \sum_{i=1}^{N} m_i u_i +\sum_{\mu=1}^{M} f_\mu v_\mu -\log { \int d \vec{x} d \vec{h} d \vec{\hat{h}}  e^{-\beta H[\vec{h}]+\sum_i x_i u_i +\sum_{\mu} i \hat{h}_\mu  v_\mu +\sum_\mu i\hat{h}_\mu( h_\mu(x)-h_\mu)}} \ .
\label{eq:functionalGamma}
\end{equation}
The total force acting on the particle $i$ is given by:
\begin{equation}
F_i= -\frac{d \mathcal{H}}{dx_i}= \sum_{\mu=1}^{M} \left(-h_\mu \theta(-h_\mu) \right) \frac{d h_\mu}{d x_i}= \sum_{\mu \in \mathcal{C}} f_\mu S_{\mu i} \ ,
\end{equation}
where $f_\mu$ is the \emph{contact force} and $S_{\mu i}=d h_\mu / d x_i$ is usually called \emph{dynamical matrix} \cite{FPSUZ}. The notation $\mu \in \mathcal{C}$ stands for those contacts such that $h_\mu <0$, namely the set of unsatisfied constraints. 
Starting from this definition of the contact force and looking at the derivative of the functional $\Gamma$ with respect to the gap, we recover:
\begin{equation}
\frac{d \Gamma(\vec{m}, \vec{f})}{d h_\mu}= \frac{d}{d h_\mu}\left( \frac{\beta}{2} \sum_{\mu=1}^{M} h_\mu^2 \theta(-h_\mu) \right) + \langle i \hat{h}_\mu \rangle \ .
\end{equation}
As we are interested in the SAT regime where the gaps are positive-definite, the only surviving term is the ensemble average value $\langle i \hat{h}_\mu \rangle$, defined above as the \emph{generalized force} $f_\mu$. The latter is obtained by differentiating the free energy with respect to the gap rather than to the actual position. 

Similarly to a spin-glass model where the free energy is a function of the overlap value, here the free energy functional depends on the \emph{self-overlap} between two particle configurations, a.k.a the Edwards-Anderson parameter, as well as on the first two moments of the forces:
\begin{equation}
q=\frac{1}{N}\sum_{i=1}^{N} m_i^2 \ , \hspace{0.2cm} 
r=-\frac{1}{\alpha N} \sum_{\mu=1}^{M} f_{\mu}^2  \ ,\hspace{0.3cm}
\tilde{r}=\frac{1}{\alpha N} \sum_{\mu=1}^{M} \langle\hat{h}^2_{\mu} \rangle \ .
\label{definitions}
\end{equation}
Hence Eq. (\ref{eq:functionalGamma}) can be rewritten as:
\begin{equation}
\Gamma(\vec{m},\vec{f})=\sum_{i=1}^{N} m_i u_i +\sum_{\mu=1}^{M} f_\mu v_{\mu}-\log \int d \vec{x} d \vec{h} d \vec{\hat{h}}  \hspace{0.15cm} e^{{S_\eta(\vec{x},\vec{h},\vec{\hat{h}})}}  \ ,
\label{pot}
\end{equation}
\begin{equation}
S_\eta(\vec{x},\vec{h},\vec{\hat{h}})=   \sum_{i=1}^N u_i  x_i + \sum_{\mu=1}^M i v_\mu \hat{h}_\mu -\lambda \sum_{i=1}^N (x_i^2-N) -\frac{\beta}{2} \sum_{\mu=1}^{M} h_\mu^2 \theta(-h_\mu)-i\sum_{\mu=1}^{M} \hat{h}_\mu (h_\mu-\eta h_\mu(x)) -\frac{b}{2} \sum_{\mu=1}^{M} (\hat{h}_\mu^2-\alpha N \tilde{r}) \ .
\label{pot2}
\end{equation}
Note that we have introduced two additional parameters compared to Eq. (\ref{eq:functionalGamma}): $\lambda$ guarantees the correct normalization on the $N$-dimensional sphere, while $b$ enforces the second moment of $i \hat{h}_\mu$. The value of the multiplier $b$ is constrained to be $1-q$ by the saddle point equation $\frac{\partial \Gamma}{\partial \tilde{r}}=0$.
As it will be clarified in the following, we also need to fix the average value of $(i \hat{h}_\mu)^2$ to write down a closed set of equations. 

The main goal of this paper is to study the low energy phase of the perceptron model at zero-temperature. 
In the zero-temperature limit we can have two different behaviors: in the SAT phase several solutions are possible and the overlap parameter $q<1$.  Conversely, in the UNSAT phase the energy has one single minimum and the overlap parameter is always equal to one. In the following we will focus on the SAT phase in the $T \rightarrow 0$ limit. In this regime the free energy corresponds to the configurational entropy of the system as a measure of the number of microstates $\Omega(v)$ with a given volume $v$. In other terms, $S \propto \log \Omega(v)$, where $\Omega(v)= \int d \vec{x} \delta(v-W(\vec{x})) \Theta_{\text{jamm}}$, the $\Theta$ function enforcing the excluded volume constraint \cite{Edwards-O, Edwards}. 

The core of our computation lies in the definition of an auxiliary \emph{effective Hamiltonian} $\mathcal{H}_{eff}= i \eta \sum_{\mu} \hat{h}_\mu h_\mu(\vec{x})$, where $\eta$ represents the parameter in terms of which we perform a Plefka-like expansion \cite{Plefka, Georges-Yedidia}. Indeed, the original Hamiltonian in Eq. (\ref{Hamiltonians}) is zero in the SAT phase and it only contributes in forcing the particles to stay close. 
In a fully-connected systems in the large $N$ limit one can recover the mean-field predictions considering only the first two terms in the expansion. 
Higher-order terms provide systematic corrections to the mean-field approximation, relevant for short-range interacting models or finite-dimensional ones.
More precisely, the expansion in $\eta$ coincides with a diagrammatic expansion in the inverse of the dimension $1/N$. In general, we need to determine the following quantity:
\begin{equation}
\Gamma(\eta)= \sum_{n=0} \frac{1}{n!} \left .\frac{\partial^{n} \Gamma}{\partial \eta^n} \right\vert_{\eta=0} \eta^n \ ,
\end{equation}
where $\Gamma$ is the free energy functional, which, for simplicity of notation, depends here only on $\eta$.
We formally expand around $\eta=0$ and then we set $\eta=1$ without any loss of generality.
The first derivative of the functional above with respect to $\eta$ coincides with the average effective Hamiltonian evaluated in the coarse-grained values, whereas the second derivative involves both the connected part of the effective Hamiltonian and the partial derivatives of the Lagrange multipliers $u_i$ and $v_\mu$. This computation gives rise to the \emph{Onsager reaction term} in Thouless-Anderson-Palmer (TAP) formalism \cite{TAP}:
\begin{eqnarray}
\frac{\partial^2 \Gamma}{\partial \eta^2} = - \Biggl \{ \langle H_{eff}^2 \rangle -\langle H_{eff} \rangle^2  +  \Biggl\langle H_{eff}  \Biggl[ \sum_i \frac{\partial u_i}{\partial \eta} (x_i -m_i) +\sum_\mu \frac{\partial v_\mu}{\partial \eta}(i\hat{h}_\mu-f_\mu) \Biggr] \Biggr\rangle \Biggr \} \ .
\label{second_deriv}
\end{eqnarray}
The resulting expression for the potential up to the second order in $\eta$ reads:
\begin{widetext}
\begin{equation}
\begin{split}
\Gamma(\vec{m},\vec{f})=&\sum_{i=1}^N \phi(m_i)+\sum_{\mu=1}^M \Phi(f_{\mu})+\left.\frac{\partial \Gamma}{\partial \eta} \right|_{\eta=0}\eta +\left.\frac{1}{2}\frac{\partial^2 \Gamma}{\partial \eta^2} \right|_{\eta=0} \eta^2+\mathcal{O}(\eta^3)=\\ 
\approx &-\frac{N}{2}\log (1-q)+\sum_{\mu} \Phi(f_{\mu})- \sum_{i,\mu} \frac{\xi_{i}^{\mu}m_i f_{\mu}}{\sqrt{N}} +\frac{\alpha N}{2}  (\tilde{r}-r)(1-q) \ .
\end{split}
\label{eq:effective_pot}
\end{equation}
\end{widetext}
Note that, while in a fully-connected ferromagnetic model the only relevant term is the first moment, as all couplings are $O(1/N)$ and all spins are equivalent \cite{Mezard_Parisi_Virasoro}, in a disordered system as the one presented here both the first and the second moments cannot be neglected.
To obtain the last line of Eq. (\ref{eq:effective_pot}), we have simply evaluated via a saddle-point computation the integral over $\vec{x}$, corresponding to the entropy of a non-interacting system constrained on a spherical manifold. Then the term $\phi(m_i)$ turns out to be proportional to $\log(1-q)$, as expected for a spherical model. 

Moreover, factorizing the terms which depend on the Lagrange multiplier $v_\mu$ and on $\hat{h}_\mu$, ${\hat{h}_\mu}^2$ respectively, the functional $\sum_\mu \Phi(f_\mu)$ can be rewritten in a more straightforward way. Note that while the integral over $\hat{h}_\mu$ is extended over all values in $(-\infty, \infty)$, the integral over the gaps $h_\mu$ can take only positive values in the SAT phase. Since $i \hat{h}_\mu$ is a real variable by definition, namely a physical force, the integration is actually performed in the complex plane and one looks at the values of $h_\mu$ and $\hat{h}_\mu$ for which the action is stationary. Then we get:
\begin{equation}
\Phi(\vec{f})={{\min_{v}}} \left[ f v -\log H{ \left( \frac{ \sigma -v}{\sqrt{1-q}} \right)} \right] \ ,
\end{equation}
where we indicated as $H(x) \equiv \frac{1}{2} \text{Erfc} \left( \frac{x}{ \sqrt{2} }\right) $.
By differentiating the expression above with respect to $v_\mu$, we immediately get the forces $f_\mu$. For a detailed computation we refer the reader to \cite{AFP}. It is worth noticing that both the method and the results discussed here for the negative perceptron can be safely generalized to sphere systems in high dimensions, where the effective potential takes roughly the same form.

Starting from Eq. (\ref{eq:effective_pot}) one immediately finds the following stationary equations for the local quantities $m_i$ and $f_\mu$, numerically solvable in an iterative scheme:
\begin{equation}
\frac{\partial \Gamma}{\partial m_i}=0  \hspace{0.2cm} \Rightarrow \hspace{0.3cm}
m_i \left(\frac{1}{1-q}-\alpha(\tilde{r}-r)\right)= \sum_{\mu} \frac{\xi_{i}^{\mu} f_{\mu}}{\sqrt{N}} \ ,
\label{saddle-p_m}
\end{equation}
\begin{equation}
\frac{\partial \Gamma}{\partial f_{\mu}}=\Phi^{'}(f_\mu)-\sum_{i}\frac{ \xi_{i}^{\mu}m_i}{\sqrt{N}}+(1-q)f_{\mu}=0 \ .
\label{eq:stat_g}
\end{equation}
An alternative procedure consists in deriving the belief propagation equations \cite{Mezard-Montanari} for the $x_i$'s and $i \hat{h}_\mu$'s and then assuming that they can be parametrized by Gaussian distributions. 
Provided that the moments can be expanded in $1/N$, we can rewrite the belief propagation equations in terms of single site quantities associated to the nodes of a factor graph \cite{Mezard-Montanari}. This procedure leads exactly to Eqs. (\ref{saddle-p_m})-(\ref{eq:stat_g}).
In an ordinary ferromagnet the solution of the equations above is very easy to find since the couplings are known and they do not depend on the space indexes separately. In a spin glass or a generic disordered system the situation is much more complex, since the $\xi_i^{\mu}$'s are random variables whose probability distribution is the only available information. However, for an infinite range model in the $N \rightarrow \infty$ limit, it is possible to prove \cite{Mezard_Parisi_Virasoro} that only a marginal modification is needed, namely to consider an auxiliary system of $N-1$ and $M-1$ variables with the $i$-th and the $\mu$-th ones removed.
Using the notation $\sum_{i} \frac{\xi_i^{\mu} m_i}{\sqrt{N}}\equiv h_\mu(\vec{m}) +\sigma$ and recalling that $\Phi^{'}(f_\mu)=v_\mu$, we can rewrite Eq. (\ref{eq:stat_g}) as:
\begin{equation}
h_\mu(\vec{m})=v_\mu-\sigma +(1-q)f_\mu = v_\mu-\sigma -\sqrt{1-q} \frac{H' \left( \frac{\sigma-v_\mu}{\sqrt{1-q}} \right) }{ H \left( \frac{\sigma-v_\mu}{\sqrt{1-q}} \right) } \ .
\label{v-f}
\end{equation}
\vspace{0.3cm}

\begin{figure}[h]
\centering
\includegraphics[scale=0.95]{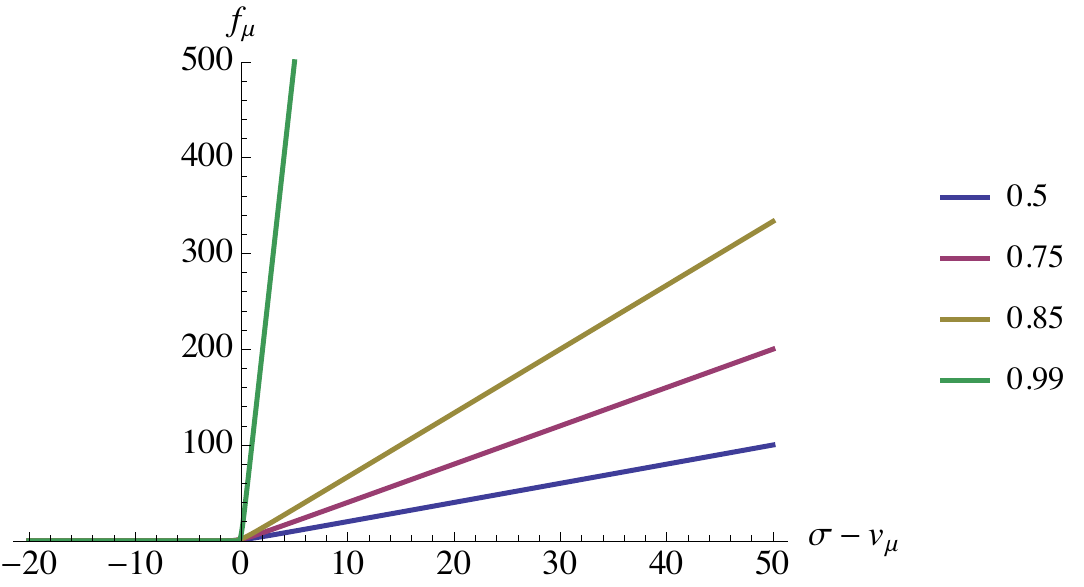}
\caption{Generalized forces as a function of $\sigma-v_\mu$ plotted for different values of the overlap $q$. In the jamming limit, as $q \rightarrow 1$, the function approaches the vertical axis (green line), in agreement with the expected divergence of the forces.}
\end{figure}
If the argument of the complementary error function $H(x)$ is much greater than one, \emph{i.e.} in the jamming limit, the last term can be simplified and the resulting expression turns out to be linear in $(\sigma-v_\mu)/\sqrt{1-q}$ with opposite sign with respect to the first piece. The two terms cancel each other out and the jamming limit is exactly identified by the condition $h_\mu \rightarrow 0$. In this regime we recover a logarithmic interaction for the effective potential as a function of the average gaps. This behavior is independent of the actual dimension of system and exactly derivable in infinite-dimensional systems \cite{AFP}.

Thus, the random gaps are written as the contribution of the so-called \emph{cavity field} in the spin-glass literature and the Onsager reaction term, the latter giving the correction with respect to the na\"{i}ve mean-field equation. This argument can be understood from Eq. (\ref{eq:stat_g}), where the value of $v_\mu$ is actually due to $m_i$ in the absence of the $\mu$-th contact. The reaction term, namely $(1-q)f_\mu$, represents instead the influence of the $\mu$-th particle on the others.
Therefore, there is a subtle difference between the effective gap $h_\mu \left( \vec{m} \right)$ and the cavity field $v_\mu-\sigma$, the field that the neighboring particles would feel if removing a single particle in the network. As mentioned above, the set of values for which $h_\mu<0$ corresponds to the effective contacts at jamming and since in the SAT phase the gaps are positive, the only possibility is to have negative values for the cavity field. 

\section{Third-order corrections to the effective potential}
In the previous Section we showed the derivation of the TAP free energy taking into account only the first two terms of the expansion, in a mean-field-like picture. 
One could be interested in defining a modified version of the perceptron - for instance a diluted model with finite-connectivity patterns $\xi^{\mu}$ - or even a finite dimensional system not exactly at jamming. In both cases, further order corrections would play a relevant role and provide a finite contribution in the perturbative expansion in the inverse of the dimension. Hence, we should take into account all the corrections to the potential coming from loopy structures by summing over triplets, quadruplets and generic combinations of links. 
For the purpose of this work the computation of the next order correction to the TAP free energy turns out to be a useful tool to understand how the coarse-grained potential deviates from its critical trend. Therefore, we need to determine the following expression \cite{Georges-Yedidia, Nakanishi-Takayama}:
\begin{equation}
\frac{\partial ^3 \Gamma}{\partial \eta^3}=  \langle H_{eff} \rangle \frac{\partial \langle H_{eff} \rangle}{\partial \eta}+ \langle H_{eff} \Upsilon_2 \rangle +  \langle H_{eff} \left( H_{eff}-\langle H_{eff} \rangle +\Upsilon_1 \right)^2 \rangle \ ,
\end{equation}
where $\Upsilon_n$ reads:
\begin{equation}
\Upsilon_n= \sum_{i} \frac{\partial }{\partial y_i} \left( \frac{\partial^n \Gamma}{\partial \eta^n} \right) (s_i -y_i)  \ .
\end{equation}
For simplicity, we indicated both derivatives, with respect to $m_i$ and $f_\mu$, as $(s_i -y_i)\frac{\partial}{\partial y_i}$.
The resulting expression for the third-order corrections is:
\begin{equation}
\begin{split}
\frac{\partial ^3 \Gamma}{\partial \eta^3} & = 
 \langle H_{eff}^3 \rangle +\langle H_{eff} \rangle \langle H_{eff}^2 \rangle -2 \langle H_{eff} \rangle ^3  -  \langle H_{eff} \rangle \alpha N r (1-q) - \langle H_{eff} \rangle \alpha N q(\tilde{r}-r) + \Biggl \langle H_{eff} \left( -\sum_{i,\mu} \frac{\delta x_i}{\sqrt{N}} \xi_i^{\mu} f_\mu \right)^2 \Biggr \rangle   \\
& + \Biggl \langle H_{eff} \left(- \sum_{i,\mu} \frac{\delta f_\mu}{\sqrt{N}} \xi_i^{\mu} m_i \right)^2  \Biggr \rangle  -  2 \Biggl \langle H_{eff}^2 \left( \sum_i \delta x_i \sum_{\mu} \frac{\xi_i^{\mu} f_\mu}{\sqrt{N}} + \sum_{\mu}\delta f_\mu \sum_i \frac{\xi_i^{\mu}m_i}{\sqrt{N}} \right) \Biggr \rangle
\label{corrections3}
\end{split}
\end{equation}
where $\delta x_i= (x_i -m_i)$ and $\delta f_\mu=(i\hat{h}_\mu-f_\mu)$ are the relative deviations of the particle positions and the contact forces from their own mean value respectively.

The first terms in Eq. (\ref{corrections3}) reminds an analogous expression for the Sherrington-Kirkpatrick (SK) model obtained from the TAP approach \cite{Mezard_Parisi_Virasoro}. The other terms are instead due to the variation of the additional parameters on which the perceptron model actually depends.
In principle, these finite-size corrections are not negligible. However, in the jamming limit, \emph{i.e.} as $q \rightarrow 1$, most of these terms can be re-expressed in a more straightforward way. 
The fourth and the fifth term respectively cancel with the next two terms with opposite sign, being their squared moments $(1-q)r$ and $(\tilde{r}-r)q$ in turn. We have to focus only on the first three terms and the very last one. As in the jamming limit the values of the positions and the forces tend to their coarse grained values, \emph{i.e.} $x_i \rightarrow m_i$ and $i \hat{h}_\mu \rightarrow f_\mu$, the last term can be neglected. The most interesting contribution comes from the first three terms. The underlying property concerning the jamming line is the presence of very weak correlations, which make a connection between the jamming transition and a mean-field-like scenario possible. This argument can be rephrased as follows:
\begin{equation}
\sum_{k,l=1}^{N} \left( \langle x_k x_l \rangle -\langle x_k \rangle \langle x_l \rangle \right)=0 \ .
\end{equation}
Thus, in the jamming limit the first three terms cancel each other out. This result is in remarkable agreement with the fact that the jamming transition is well described in terms of binary interactions only \cite{Wyart-Brito}. We shall clarify this point in the next Section, concerning the analysis of the typical scaling of the forces.

The idea supporting our computation is that all powers, except for the first two, actually vanish in the jamming limit. This means that, by considering the functional derivative $\left . \frac{\partial^{n} \Gamma}{\partial \eta^{n}} \right \vert_{\eta=0}$ evaluated at $\eta=0$, the result should be identically zero. In the lowest dimensional case, for $n=1$, this simplification is immediate:
\begin{equation}
\left . \frac{\partial \Gamma}{\partial \eta} \right \vert_{\eta=0}  =  \sum_{i=1}^{N} \frac{ \partial u_i}{\partial \eta} \left( m_i - \frac{\partial \Gamma}{ \partial u_i} \right) + \sum_{\mu=1}^{M} \frac{ \partial v_\mu}{\partial \eta} \left( f_\mu -  \frac{\partial  \Gamma}{\partial v_\mu} \right)   + \Biggr \langle \left( \sum_{i=1}^{N} \frac{ \xi_i^{\mu} x_i}{ \sqrt{N}} -\sigma \right) \Biggr \rangle_c \ .
\end{equation}
The first two terms are zero thanks to the fact that $m_i$ and $v_\mu$ are fixed by a Legendre transform of the potential $\Gamma(\vec{m}, \vec{f})$, while the last term corresponds to the connected correlation function of the average gap, which is roughly zero in the jamming limit. A more detailed computation of the fourth order term and beyond would not change the conclusion. The underlying reason is related to the isostaticity condition \footnote{At the jamming threshold the number of degrees of freedom in the system exactly equals the number of constraints, in other words the coordination number $z=z_c$. Isostaticity is also responsible for marginal mechanical stability.}.

It is worth highlighting that in glassy systems a perturbative diagrammatic expansion of the correlation functions can be established if the cage of the particles is sufficiently small \cite{Mezard-Parisi, Parisi-Slanina}, namely in the high pressure regime. The hypernetted chain (HNC) approximation \cite{Hansen-McD, DeDominicis-Martin} does not work for small cage radius, but alternative approaches can be exploited. In particular, in \cite{Parisi-Zamponi2010} the authors proposed a method that allows writing the correlation functions of the glass as the correlation functions of the effective liquid. 
In that case, the contributions due to three-point correlators can be factorized and rewritten as a function of two-point correlators only. Our result, based on the determination of a well-defined potential exclusively in terms of the first two moments in the jamming limit, seems to be directly correlated to this issue.

\section{Leading and subleading contributions to the forces near jamming}
\label{sec:forces}
The experimental determination of inter-particle forces in glassy materials is generally a complicated task. Conversely, from an analytical point of view, the distribution of forces can be exactly reconstructed, at least in the jamming limit. In this Section we show the connection between the effective forces and the gaps and we determine their leading and subleading contributions. The computation is performed in the perceptron model where only one \emph{annealed} particle interacts with a quenched background of spherical obstacles. Anyway, the generalization to sphere systems is immediate: in that case the gaps will depend on two labels $\alpha\beta$ identifying the two interacting particles. 

The main difficulty in determining the effective interactions in amorphous systems stems from the impossibility of writing down a simple relation between the force and the gap as soon as one attempts to extend the formalism beyond jamming. Indeed, upon decreasing the density, there is no reason to believe that the effective forces should remain binary. 
As we briefly discussed in the Introduction, two kinds of corrections should emerge in the expression of the generalized force, one related to finite-size effects and another due to the increasing distance from jamming. Let us focus on the first type. As in the jamming limit Eq. (\ref{corrections3}) reduces to zero, its derivatives with respect to $f_\mu$ turn out to be trivially zero, confirming the starting hypothesis that Eq. (\ref{eq:stat_g}) still holds. 

We now investigate how the mutual relation between the forces and the gaps, even in a mean-field-like scenario, is modified supposing to increase the distance from jamming. 
The starting point is the definition of the potential $\Phi(\vec{f})$:
\begin{equation}
\Phi(\vec{f}) =   {{\min_{v}}} \left[f  v -\log H{ \left( \frac{\sigma- v}{\sqrt{1-q}} \right)} \right]  \approx {{\min_{v}}}\Biggl \{ f \cdot v +\theta(\sigma-v) \Biggl [ \frac{(\sigma-v)^2}{2(1-q)}  + \log \left( \frac{\sigma -v}{\sqrt{1-q}} \right) \Biggr ]  \Biggr \} \ .
\label{Phi}
\end{equation}
We aim to refine our estimate with respect to the previous expectation \cite{AFP} by taking into account also the subleading terms in the asymptotic expansion of the complementary error function, provided that $(1-q)$ approaches zero in the jamming limit. 
Inserting the following expression in the potential:
\begin{equation}
H(x)=  \frac{1}{2} \text{Erfc}\left( \frac{x}{\sqrt{2}} \right) 
\approx \frac{e^{-x^2/2}}{\sqrt{2\pi} x} \left( 1+ \sum_{n=1}^{\infty} (-1)^n \frac{(2n)!}{n!(\sqrt{2} x)^{2n}} \right) \ ,
\end{equation}  
we get:
\begin{equation}
\Phi(\vec{f})  \approx  {{\min_{v}}} \Biggl \{ f \cdot v +\theta(\sigma-v) \Biggl[ \frac{(\sigma-v)^2}{2(1-q)} +\log \left( \frac{\sigma -v}{\sqrt{1-q}}\right)   -\log {\left( 1- \frac{1}{[(\sigma- v)/\sqrt{1-q})]^2 }\right) } \Biggr] \Biggr \} \ .
\end{equation}
Differentiating with respect to $v$ as before, we get a better approximation for the generalized force:
\begin{equation}
f= \frac{\sigma-v}{1-q} +\frac{1}{\sigma-v} -\frac{2(1-q)}{(\sigma-v)^3 \left[ 1- \frac{(1-q)}{(\sigma-v)^2} \right] }  \ .
\label{force}
\end{equation}
The jamming limit corresponds to $\frac{\sigma-v}{\sqrt{1-q}} \gg 1$, from which one immediately notices that the gaps tend to zero at jamming. In the opposite case, namely for $\frac{\sigma-v}{\sqrt{1-q}} \ll 1$, the contribution due to the logarithmic term in Eq. (\ref{Phi}) vanishes, as correctly expected by the definition of the error function.

Assuming that $q$ is not exactly one, but very close to it, we can expand the last term as a sum of odd powers of  $\sigma-v$, which leads to:
\begin{equation}
\begin{split}
f_\mu  & \approx \left( \frac{\sigma-v_\mu} {1-q} \right) \Biggr[ 1+ \frac{1-q}{(\sigma-v_\mu)^2}-\frac{2(1-q)^2}{(\sigma-v_\mu)^4}  -\frac{2(1-q)^3}{(\sigma-v_\mu)^6}+.... \Biggr] \\
& \approx \frac{1}{h_\mu(\vec{m})} \left[ 1+ \frac{ h_\mu(\vec{m})^2}{1-q}-\frac{2 h_\mu(\vec{m})^4}{(1-q)^2}+... \right] \\ \\
 f_\mu & \approx \frac{1}{h_\mu(\vec{m})} \mathcal{G} \left( \frac{h_\mu(\vec{m})}{\sqrt{1-q}} \right) \ .
\label{force_scaling}
\end{split}
\end{equation}
The intermediate expression in Eq. (\ref{force_scaling}) is justified by the fact that at the leading order the term $\frac{\sigma-v}{1-q}$ coincides with the inverse gap, as largely explained in \cite{AFP} and expected near jamming. Note that this scaling is valid only when approaching the critical transition from the SAT phase. The subsequent terms, including odd powers of $\frac{1}{\sigma-v}$, seem to encode the effect of a rescaled inverse pressure, which typically vanishes in the SAT region. More specifically, the logarithmic interaction near jamming emerges once that terms of order $h_\mu^2/(1-q)$ are neglected. They instead would contribute in a different regime of the phase diagram \cite{AFP}. This result has been proven for a fully-connected system in the thermodynamic limit, in remarkable agreement with the argument proposed by Wyart \emph{et al.} \cite{Wyart, Wyart-Brito, DeGiuli-Wyart} for three-dimensional hard-sphere glasses. It supports the idea of a kind of universal behavior at jamming, independent of the dimension.

The last line of Eq. (\ref{force_scaling}) can be easily understood by looking at the connected part of the average gap:
\begin{equation}
\langle h_\mu^2 \rangle_c = \frac{1}{N}\sum_{ij}\xi_i^{\mu} \xi_j^{\mu} \left(\langle x_i x_j \rangle -m_i m_j  \right) =1-q \ .
\end{equation}
Given this relation, Eq. (\ref{force_scaling}) can be written in a more compact way in terms of a scaling function $\mathcal{G}$:
\begin{equation}
f_\mu \approx \frac{1}{\langle h_\mu \rangle } \mathcal{G} \left( \frac{\langle h_\mu \rangle}{{\langle h_\mu^2 \rangle_c}^{1/2}}, \alpha \right) \ ,
\end{equation}
which gives rise to two different behaviors depending on the specific limit. As $\alpha \rightarrow \alpha_J$ the scaling function $\mathcal{G} \rightarrow 1$, confirming that the only relevant scale is the inter-particle gap, whereas for $\alpha < \alpha_J$ a full expression for $\mathcal{G}$ is needed.
Indeed the crossover regime, determining where the logarithmic potential is no longer valid, is given by the condition: $h_\mu  \sim \sqrt{1-q}$. 
Another way to clarify this point is to consider directly the expression for the generalized forces $f_\mu$, which reads:
\begin{equation}
f_\mu= -\frac{1}{\sqrt{1-q}} \frac{H' \left( \frac{\sigma-v_\mu}{\sqrt{1-q}} \right) }{ H \left( \frac{\sigma-v_\mu}{\sqrt{1-q}} \right) } \ .
\label{forces-}
\end{equation}
Eq. (\ref{v-f}), reported below, highlights a direct connection between the forces and the gaps:
\begin{equation}
h_\mu(\vec{m})=v_\mu-\sigma +(1-q)f_\mu  \ .
\end{equation}
Inserting the equation above in (\ref{forces-}), we get:
\begin{equation}
\sqrt{1-q} f_\mu= -\frac{H' \left( \frac{ -h_\mu}{\sqrt{1-q}}+\sqrt{1-q}f_\mu \right) }{ H \left( \frac{ -h_\mu}{\sqrt{1-q}}+\sqrt{1-q}f_\mu \right) } \ .
\label{inversion_f}
\end{equation}
By inverting this function with respect to $f_\mu$ we can immediately obtain the typical trend shown in Fig.(\ref{fig:scalingfunction}), divergent as the gaps shrink to zero and finite otherwise.
\vspace{0.4cm}

\begin{figure}[h]
\centering
\begin{subfigure}
\centering
\includegraphics[scale=0.68]{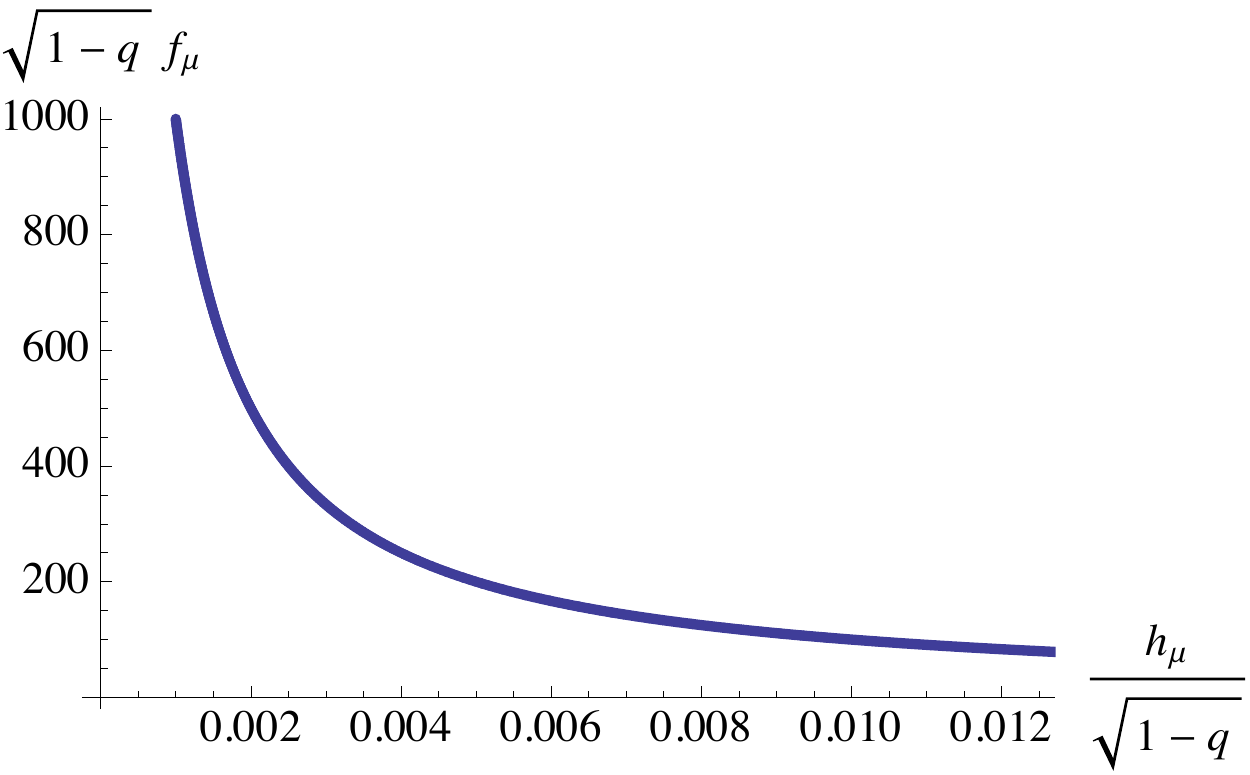}
\label{fig:scalingfunction}
\end{subfigure}
\hfill
\begin{subfigure}
\centering
\includegraphics[scale=0.68]{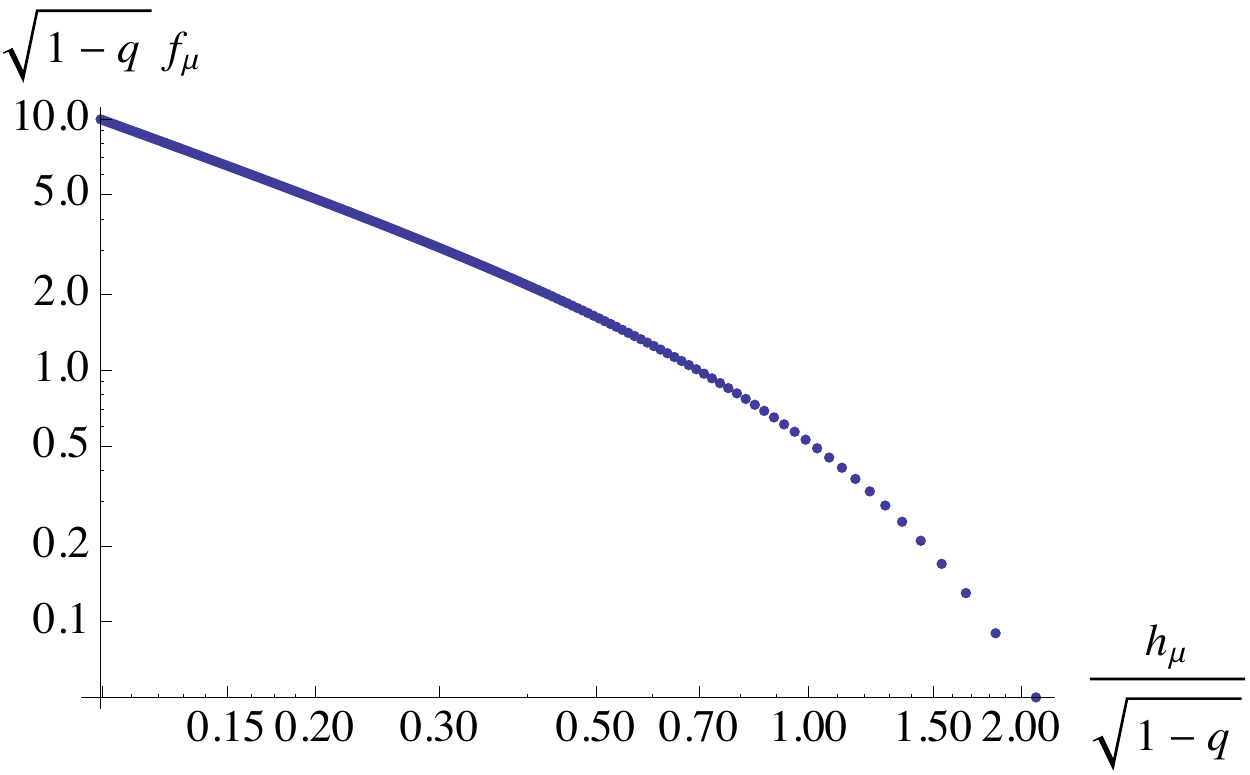}
\end{subfigure}
\caption{Scaling function showing the generalized forces as a function of the gaps, in linear and log-log scales, well fitted via a power law $a+b \left(\frac{h_\mu}{\sqrt{1-q}}\right)^{-c}$ in the small-gap regime. }
\label{fig:scalingfunction}
\end{figure}
Eq. (\ref{force_scaling}) and consequently Eq. (\ref{inversion_f}) suggest a deep analogy between the free energy of a hard-sphere glass and the energy of an athermal network of logarithmic springs \cite{Wyart, Wyart-Brito}, when looking at the dynamics on a time interval much greater than the collisional time but smaller than the structural relaxation time. This leads to the determination of a contact network and, thanks to the fact that all configurations are equiprobable at jamming, a one to one mapping between the particle displacements and the gaps can be established. The total number of contacts equals the number of degrees of freedom, according to the isostaticity condition. In this case, a simple relationship between the forces and the gaps can be determined as well. 

One might wonder why this relation should be valid in dimensions higher than one, where the mapping is no longer linear: the answer again lies in the underlying isostaticity condition, which characterizes the jamming transition.
However, upon increasing the distance from the jamming line this condition does not hold and the forces are not only functions of $h_\mu$ but of a complex combination of random parameters.
No analytical predictions about the typical scaling of the forces taking into account also subleading terms are available. Several numerical simulations have been carried out attempting to explain the observed behavior \cite{Wyart, Procaccia_etal2016}. In particular, in \cite{Wyart, Wyart1} the authors showed that the deviation of the force from its leading behavior can be estimated numerically in molecular dynamics and the subleading term should be of order of the number of effective contacts $\delta z=z-z_c$. 

Another related issue concerns the distribution of the effective forces acting on a single particle, instead of the total force distribution that is well-known to scale near the jamming line as $P(f) \sim f^{\theta}$ with $\theta=0.42311$. This computation can be presumably done in the cavity formalism \cite{Mezard-Montanari} suggesting another interesting direction for upcoming research.

\section{Scaling behaviors and crossover regime}
In Section (\ref{sec:forces}) we have discussed the leading behavior of the forces near jamming and the emergence of a smooth logarithmic interaction in the perceptron model, computable for hard-sphere systems in high dimension as well. However, as $h_\mu \sim \sqrt{1-q}$ the transition towards a logarithmic regime is progressively smeared out. 

In this Section we focus on the scaling functions describing the SAT and the UNSAT phase. In particular, our aim is understading their matching in the crossover region.
We shall exploit the main predictions of the full RSB solution largely explained in \cite{FPSUZ} in order to determine the crossover temperature-dependent behavior between these two phases.

At low temperature in the UNSAT phase the overlap has a simple dependence on temperature given by:
\begin{equation}
1-q= \chi T + O(T^2) \ ,
\end{equation}  
where $\chi$ is determined by the condition:
\begin{equation}
\left( 1+ \frac{1}{\chi} \right)= 
\sqrt{\frac{\alpha}{\alpha_J(\sigma)}} \ .
\end{equation}
The parameter $\alpha_J(\sigma)$ is the critical value on the jamming line. A generic expression for it was derived first by Gardner in the convex perceptron \cite{Gardner}.

The zero-temperature limit should be carefully performed, sending $T$ and $1-q= \chi T$ to zero simultaneously.
At jamming $\chi \rightarrow \infty$ and $q \rightarrow 1$, which determines two different scaling solutions depending on the values of $q$ and $q^{*}$, the Edwards-Anderson parameter and the matching point respectively. The matching point corresponds to the condition $\chi P(1,0) \sqrt{1-q^{*}} \sim 1$, where the probability distribution $P(q,h)$ is evaluated in $q=1$ and $h=0$ and it verifies the Parisi equation \cite{Mezard_Parisi_Virasoro}. If $q \gg q^{*}$ we recover the ordinary UNSAT phase, while for $q \ll q^{*}$ the jamming solution occurs. We know that in the UNSAT phase the pressure is proportional to the first moment of the gap $[h]$, which in turn satisfies the following relation $[h] \equiv 1/N \sum_{\mu=1}^{M} h_\mu \theta(-h_\mu) \propto 1/\chi^{2}$. Using these relations we have \cite{FPSUZ}:
\begin{equation}
(1-q^{*}) \sim \chi^{\frac{k}{1-k}} \ ,
\end{equation}
with an exponent $k \approx 1.41$. To make progress, we note that close to jamming the fullRSB equations show a scaling regime.
We focus on the regime in which the Edwards-Anderson parameter is close to the cut-off value, $q \sim q^{*}$, and we deduce the typical behavior in temperature. Let us suppose that the temperature is raised by a finite amount, yielding:
\begin{equation}
(1-q^{*}) \sim \chi^{\frac{k}{1-k}} \sim \chi T .
\end{equation}
From this relation we also conclude that:
\begin{equation}
T^{*} \sim \chi^{\frac{2k-1}{1-k}} \ .
\end{equation}
Given that in the soft-sphere regime (UNSAT) the pressure scales as $p \sim 1/\chi^2$ \cite{FPUZ, FPSUZ}, we get:
\begin{equation}
T^{*} \sim p^{\frac{2k-1}{2k-2}} \ .
\label{T-p}
\end{equation}
For $T \sim T^{*}$ the UNSAT phase and the jamming solution cannot be distinguished. Note that the relation (\ref{T-p}), connecting temperature and pressure, exactly coincides with the one proposed in \cite{DeGiuli-Wyart} based on an Effective Medium Theory argument. 

Moreover, in the SAT phase $ (1-q) \sim \epsilon^{k}$, where $\epsilon$ stands for the linear distance from the jamming line. These two relations together lead to the condition:
\begin{equation}
T^{*} \sim \epsilon^{2k-1} \ .
\end{equation}
Under these assumptions we should be able to define a scaling function of the form:
\begin{equation}
(1-q) \sim \epsilon^{k} \mathcal{F} \left( T \epsilon^{1-2k} \right) \ ,
\end{equation}
which guarantees the correct trend in each regime, either when its argument diverges or goes to zero. According to this simple argument, three different regimes can be highlighted: a HS/SAT regime, characterized by a zero energy manifold and studied in this paper by means of the TAP formalism, a SS/UNSAT regime, whose low-energy vibrational properties have been largely analyzed in \cite{FPUZ}, and an \emph{anharmonic} regime signaled by the crossover temperature $T^{*}$, which can be also related to the linear distance $\epsilon$ from jamming. Below $T^{*}$ the system actually consists of an assembly of soft harmonic particles, whereas above it its vibrational properties turn out to be indistinguishable from those of a hard-sphere system. 

\section{Conclusions}
We have presented a simple model of continuous constraint satisfaction problem (CSP), the negative perceptron, which displays a critical jamming transition. According to whether the constraints are violated or not, this model displays two different phases: a SAT phase, corresponding to a hard-sphere (HS) regime, and an UNSAT one, which can be mapped to a soft-sphere (SS) problem, well-described by a harmonic potential in the average gaps. This SAT/UNSAT transition exactly coincides with the jamming line. 

Our main goal is to capture the most relevant features in the HS regime and to specialize then the analysis to the jammed phase.
In line with the derivation proposed in \cite{AFP} of the TAP free energy, which serves as a coarse-grained functional after integrating out fast degrees of freedom, we have developed here the computation up to the third order. The analytical scheme is based on a formal Plefka-like expansion of the free energy, valid both in the high-temperature phase and in the low-temperature one. The results obtained for the negative perceptron can be safely generalized to high-dimensional sphere models allowing to get to the same conclusions.

Our analysis shows that higher order corrections do not contribute in the jamming limit, as correctly expected according to the isostaticity argument, a very general argument independent of the dimension of the system. These results suggest the idea of a close link between the jamming regime and a mean field scenario. 
Conversely, third and higher order corrections turn out to be relevant in accounting for finite-dimensional systems not exactly at jamming. They can be of great interest for numerical simulations and real glasses.

From the analysis of the effective potential near the jamming line, we have also derived the leading and subleading contributions in the expression of the contact forces, which correctly diverge at jamming and are finite away from the critical line. 
The subleading contributions can be embedded in a scaling function depending on the average gaps and the distance from jamming. The behavior of the scaling function, bridging contact forces with effective gaps, has been analyzed in this framework.

The discussion about the typical scaling laws dominating the jamming phase naturally leads to the investigation of a crossover regime between the SAT and the UNSAT phase of the perceptron model. We have determined a crossover temperature and connected it to other physical quantities of the model, such as the pressure and the linear distance from the jamming line. This crossover temperature plays a central role as it defines two different regimes: below that, the system behaves like a zero-temperature assembly of soft particles, otherwise it enters the entropic-like regime.

\section*{Acknowledgments}
I would like to warmly thank G. Parisi and S. Franz to whom I am indebted for their precious support and advice during my PhD. 

I also thank E. DeGiuli, P. Urbani, P. Vivo for useful discussions and for a critical reading of the manuscript. 
I acknowledge the Physics Department of the University of Rome Sapienza and the LPTMS of the University Paris-Sud, where part of this work has been done. 

This work was supported by grants from the Università Italo-Francese / Universit\'e Franco-Italienne (C2-20, A. Altieri) and from the Simons Foundation (No. 454941, Silvio Franz; No. 454949, Giorgio Parisi).

\vspace{0.5cm}

\begin{widetext}
\appendix
\section{Detailed computation of the third-order corrections to the free energy}
In this Appendix we present a detailed derivation of the third-order term in the TAP free energy, exploiting a Plefka-like expansion \cite{Plefka, Georges-Yedidia, Nakanishi-Takayama}. We have to evaluate the following expression:
\begin{equation}
\frac{\partial ^3 \Gamma}{\partial \eta^3}= \langle H_{eff} \rangle \frac{\partial \langle H_{eff} \rangle}{\partial \eta}+\langle H_{eff} \Upsilon_2 \rangle + \langle H_{eff} \left( H_{eff}-\langle H_{eff} \rangle +\Upsilon_1 \right)^2 \rangle \ ,
\label{third-ord}
\end{equation}
where $\Upsilon_n$ reads:
\begin{equation}
\Upsilon_n= \sum_{i} \frac{\partial }{\partial y_i} \left( \frac{\partial^n \Gamma}{\partial \eta^n} \right) (s_i -y_i)  
\end{equation}
and $\langle H_{eff} \rangle =\sum \limits_{i,\mu} \frac{\xi_i^{\mu} m_i} {\sqrt{N}}f_\mu $.
We can rewrite Eq. (\ref{third-ord}) as:
\begin{equation}
\begin{split}
\frac{\partial ^3 \Gamma}{\partial \eta^3}=& -\langle H_{eff} \rangle \frac{\partial ^2 \Gamma}{ \partial \eta^2} +\langle H_{eff} \Upsilon_2 \rangle+ \langle H_{eff} \left( H_{eff}- \langle H_{eff} \rangle +\Upsilon_1 \right)^2 \rangle= \\
=& \langle H_{eff} \rangle \left[ \langle H_{eff}^2 \rangle - \langle H_{eff} \rangle^2 -\langle H_{eff} \sum_{i} (s_i-y_i) \frac{\partial \langle H_{eff} \rangle }{\partial y_i}\rangle  \right]+ \Biggl \langle H_{eff} \sum_i (s_i-y_i) \frac{\partial}{\partial y_i} \frac{\partial ^2 \Gamma}{\partial \eta^2} \Biggr \rangle + \\
+& \Biggl \langle H_{eff} \left( H_{eff}-\langle H_{eff} \rangle - \sum_{i} \frac{\partial \langle H_{eff} \rangle }{\partial y_i}  (s_i-y_i) \right)^2 \Biggr \rangle 
\end{split}
\end{equation}
Expanding the last square term and differentiating explicitly with respect to $m_i$ and $f_\mu$, we get:
\begin{equation}
\begin{split}
\frac{\partial ^3 \Gamma}{\partial \eta^3}=& \langle H_{eff}^3\rangle+\langle H_{eff} \rangle \langle H_{eff}^2 \rangle  -2 \langle H_{eff} \rangle^3 + \langle H_{eff} \rangle \langle H_{eff} \sum_i  \frac{\partial \langle H_{eff} \rangle} {\partial m_i} (x_i-m_i) \rangle +\\
+& \langle H_{eff} \rangle \langle H_{eff} \sum_\mu \frac{\partial \langle H_{eff} \rangle} {\partial f_\mu} (i\hat{h}_\mu -f_\mu)\rangle + \Biggl \langle H_{eff} \left( \sum_i  \frac{\partial}{\partial m_i} \frac{\partial\Gamma}{\partial \eta} (x_i-m_i) \right)^2 \Biggr \rangle  +\Biggl \langle H_{eff} \left( \sum_\mu \frac{\partial}{\partial f_\mu} \frac{\partial\Gamma}{\partial \eta} (i\hat{h}_\mu-f_\mu) \right)^2 \Biggr \rangle +\\
-& 2 \Biggl \langle H_{eff}^2  \left[ \sum_i (x_i-m_i) \frac{\partial \langle H \rangle}{\partial m_i} \rangle +\sum_\mu (i\hat{h}_\mu -f_\mu) \frac{\partial \langle H_{eff} \rangle} {\partial f_\mu} \right] \Biggr \rangle= \\
=& \langle H_{eff}^3 \rangle +\langle H_{eff} \rangle \langle H_{eff}^2 \rangle  -2 \langle H_{eff} \rangle^3 +\langle H_{eff} \rangle \langle \sum_{ij, \mu \nu} \frac{\xi^{\mu}_i \xi^{\nu}_j}{N} x_i (x_j -m_j) i\hat{h}_\mu f_\nu \rangle + \\
+&  \langle H_{eff} \rangle \langle \sum_{ij, \mu \nu} \frac{\xi_{i}^{\mu} \xi^{\nu}_{j}}{N} x_i m_j i \hat{h}_\mu (i \hat{h}_\nu-f_\nu) \rangle + \Biggl \langle H_{eff} \left( \sum_i (x_i - m_i) \sum_{\mu} \left(-\frac{\xi_{i}^{\mu} f_\mu}{\sqrt{N}} \right) \right)^2 \Biggr \rangle +\\
+&  \Biggl \langle H_{eff} \left( \sum_\mu (i\hat{h}_\mu-f_\mu) \sum_{i} \left(- \frac{\xi_i^{\mu} m_i}{\sqrt{N}} \right) \right)^2 \Biggr \rangle - 2 \Biggl \langle H_{eff}^2 \left[ \sum_i (x_i -m_i) \sum_{\mu} \frac{\xi_i^{\mu} f_\mu}{\sqrt{N}} \Biggr\rangle + \sum_{\mu} (i\hat{h}_\mu-f_\mu) \sum_i \frac{\xi_i^{\mu}m_i}{\sqrt{N}} \right] \Biggr \rangle 
\end{split}
\end{equation}
Off-diagonal terms do not contribute in the computation, \emph{i.e.} we have to consider only diagonal terms, with $i=j$ and $\mu=\nu$. For more details, we refer the reader to the Appendix of \cite{AFP}. The final expression reduces to:
\begin{equation}
\begin{split}
\frac{\partial ^3 \Gamma}{\partial \eta^3} = 
& \langle H_{eff}^3 \rangle +\langle H_{eff} \rangle \langle H_{eff}^2 \rangle -2 \langle H_{eff} \rangle ^3 - \langle H_{eff} \rangle \alpha N r (1-q) - \langle H_{eff} \rangle \alpha N q(\tilde{r}-r)+ \Biggl \langle H_{eff} \left( -\sum_{i,\mu} \frac{\delta x_i}{\sqrt{N}} \xi_i^{\mu} f_\mu \right)^2 \Biggr \rangle   \\
& + \Biggl \langle H_{eff} \left(- \sum_{i,\mu} \frac{\delta f_\mu}{\sqrt{N}} \xi_i^{\mu} m_i \right)^2  \Biggr \rangle  - 2 \Biggl \langle H_{eff}^2 \left( \sum_i \delta x_i \sum_{\mu} \frac{\xi_i^{\mu} f_\mu}{\sqrt{N}} + \sum_{\mu}\delta f_\mu \sum_i \frac{\xi_i^{\mu}m_i}{\sqrt{N}} \right) \Biggr \rangle \ .
\end{split}
\end{equation}
\end{widetext}

\end{document}